\documentclass[12pt]{article}

\usepackage{amssymb}
\usepackage{times}
\usepackage{graphicx}
\usepackage{enumerate}
\usepackage{amsmath}
\usepackage{bm}
\usepackage{eucal}
\usepackage{url}
\usepackage{xcolor}
\usepackage{calc}
\usepackage{array}
\usepackage{indentfirst}
\usepackage{fancyhdr, calc, lastpage}       % headers/footers

\makeatletter
\def\@seccntformat#1{\csname the#1\endcsname. }

\usepackage[paperwidth=20.5cm,paperheight=29.7cm,width=18cm,height=25.5cm,top=24mm,left=14mm,headsep=11mm]{geometry}

\newcommand{\I}{\mathrm{i}}
\newcommand{\E}{\mathrm{e}}

\newcommand{\im}{\mathop{\mathrm{Im}}}

\newcommand\setheader[2]{
    \fancyhead[L]{\footnotesize #1}
    \fancyhead[R]{\footnotesize #2}
}

\renewcommand\title[1]{{\linespread{1} \noindent\LARGE \bf \hskip2.25pc \parbox{.8\textwidth}{%
\LARGE \bf \begin{center} #1 \end{center}\rm }
\rm\normalfont\normalsize} }
\renewcommand\author[1]{{\linespread{1} \noindent\hskip2.25pc \parbox{.8\textwidth}{%
   \normalsize \bf \begin{center} #1 \end{center}\rm } \vskip-1.4pc }}
\newcommand\address[1]{{\linespread{1} \noindent\hskip2.25pc \parbox{.8\textwidth}{%
   \footnotesize \it \begin{center} #1 \end{center}\rm }  \normalsize \vskip-1pc }}
\newcommand\email[1]{\vskip-.3cm \noindent\parskip0pc\hskip2.25pc \footnotesize%
   \parbox{.8\textwidth}{\begin{center}\it #1 \rm \end{center} } \normalsize  \vskip-.2cm}
\newcommand\PACS[1]{\vskip-2.75pc \begin{center}\parbox{.8\textwidth}{\small\bf PACS numbers: \rm #1 \hfill} \end{center}\vskip4pt}%

\renewenvironment{abstract}%%
{\vskip1pc\noindent\begin{center} \begin{minipage}{.8\textwidth} {\bf Abstract: } }
{ \vspace{.25cm} \end{minipage}\end{center}\normalsize\vskip-1.5pc}%

\def\fps@table{h}%\def\fps@table{!htb}

\makeatletter
\renewcommand\@seccntformat[1]{\csname the#1\endcsname.\hspace{.1cm}}

\renewcommand\section{\@startsection {section}{1}{0pt}%
                                     {-2ex plus -1ex minus -.2ex}%
                                     {0.65ex plus 1.2ex}%
                                     {\normalsize\bfseries}}
\renewcommand\subsection{\@startsection{subsection}{2}{0pt}%
                                     {-2.25ex plus -1ex minus -.2ex}%
                                     {.45ex plus .2ex}%
                                     {\normalsize\itshape}}

\begin{document}

\setheader{Mode Gaussian beam tracing}{Trofimov et al}

\title{Mode Gaussian beam tracing}

\author{Mikhail Yu. Trofimov, Alena D. Zakharenko, Sergey B. Kozitskiy}
\address{V.I.Il`ichev Pacific Oceanological Institute, 43, Baltiyskaya Street,\\ Vladivostok, 690041, Russia}
\email{trofimov@poi.dvo.ru, zakharenko@poi.dvo.ru, skozi@poi.dvo.ru}

% ================================================================================================
\begin{abstract}
An adiabatic mode Helmholtz equation for 3D underwater sound
propagation is developed. The Gaussian beam tracing in this case is
constructed. The test calculations are carried out for the
crosswedge benchmark and proved an excellent agreement with the
source images method.
\end{abstract}

\section{Introduction}

The problem of sound propagation across the slope in three dimensions is considered by the
method of summation of mode Gaussian beams~\cite{cerpop,porter}.
In our case, no interaction of modes is necessary to model correctly across slope propagation.
The paper is organized as follows. After formulation of the problem in
section 2, we consider an adiabatic mode Helmholtz equation and the corresponding
parabolic equation in the ray-centered coordinates.
In the next section we develop certain details related to the mode Gaussian beams propagation.
After that we illustrate the efficiency of the obtained equation by the
numerical simulation of sound propagation for the standard ASA wedge
benchmark, as it was performed in the paper~\cite{lin} %,sturm}
for the case of the 3D parabolic equation.
The paper ends with a brief conclusion.

\section{Basic Equations and Boundary Conditions}

We consider the propagation of time-harmonic sound in the
three-dimen\-si\-onal waveguide $$\Omega=\{(x,y,z)| 0 \leq x\leq
\infty, -\infty \leq y\leq \infty,  -H\leq z \leq 0 \}$$
 (the $z$-axis is directed upward),
described by the acoustic Helmholtz equation
\begin{equation} \label{Helm}
\left(\gamma P_x\right)_x + \left(\gamma P_y\right)_y
 + \left(\gamma P_z\right)_z + \gamma\kappa^2 P = 0\,,
\end{equation}
where $\gamma = 1/\rho$, $\rho=\rho(x,y,z)$ is the density, $\kappa$ is the wave-number.
We assume the appropriate radiation conditions at infinity in the $x,y$ plane, the pressure-release
boundary condition $P=0$ at $z=0$
and the rigid boundary condition $\partial P/\partial z=0$ at $z= -H$.
The parameters of the medium can be discontinuous at the nonintersecting smooth interfaces  $z=h_1(x,y),\ldots,h_m(x,y)$,
where the usual interface conditions\sloppy
\begin{equation}\label{InterfCond}
\begin{split}
& P_+ = P_-\,,\\
&\gamma_+\left(\frac{\partial P}{\partial z}-h_x\frac{\partial P}{\partial x}-
h_y\frac{\partial P}{\partial y}\right)_+ =
\gamma_-\left(\frac{\partial P}{\partial z}-h_x\frac{\partial P}{\partial x}-
h_y\frac{\partial P}{\partial y}\right)_-\,,
\end{split}
\end{equation}
are imposed. Hereafter, we use the denotations $f(z_0,x,y)_+=\lim_{z\downarrow z_0}f(z,x,y)$ and
$f(z_0,x,y)_-=\lim_{z\uparrow z_0}f(z,x,y)$.
As will be seen below, it is sufficient to consider the case $m=1$, so we set $m=1$ and denote $h_1$ by $h$.
\par
We introduce a small parameter $\epsilon$ (the ratio
of the typical wavelength to the typical size of medium inhomogeneities), the slow
variables $X=\epsilon x$ and $Y=\epsilon y$ and the fast variables $\eta=(1/\epsilon) \Theta(X,Y)$ and $\xi=(1/\sqrt{\epsilon})\Psi(X,Y)$
and postulate the following
expansions for the acoustic pressure $P$ and the parameters $\kappa^2$, $\gamma$ and $h$:
\begin{equation}\label{expans}
\begin{split}
 & P=P_0(X,Y,z,\eta,\xi)+\sqrt{\epsilon} P_{1/2}(X,Y,z,\eta,\xi)+\cdots\,,\\
 & \kappa^2 = n_0^2(X,Y,z) + \epsilon\nu(X,Y,z,\xi)\,, \\
 & \gamma  =  \gamma_0(X,Y,z) + \epsilon\gamma_1(X,Y,z,\xi)\,,\\
 & h = h_0(X,Y) + \epsilon h_1(X,Y,\xi)\,.
\end{split}
\end{equation}
To model attenuation effects, we admit $\nu$ to be complex. Namely, we take $\im\nu = 2\mu\beta n_0^2\,,$ where
$\mu = (40\pi\log_{10}e)^{-1}$ and $\beta$ is the attenuation  in decibels per wavelength.

Following the generalized multiple-scale method~\cite{nay}, we replace derivatives in equation~(\ref{Helm}) by the rules
\begin{equation*}
\begin{split}
& \frac{\partial}{\partial x} \rightarrow \epsilon\left(\frac{\partial}{\partial X}+\frac{1}{\sqrt{\epsilon}}\Psi_X\frac{\partial}{\partial \xi}
+\frac{1}{\epsilon}\Theta_X\frac{\partial}{\partial \eta}\right)\,,\\
& \frac{\partial}{\partial y} \rightarrow \epsilon\left(\frac{\partial}{\partial Y}\,+\frac{1}{\sqrt{\epsilon}}\Psi_Y\frac{\partial}{\partial \xi}
+\frac{1}{\epsilon}\Theta_Y\frac{\partial}{\partial \eta}\right)\,.
\end{split}
\end{equation*}
 Given the postulated expansions, the equation under consideration becomes
\begin{equation} \label{2}
\begin{split}
&\epsilon^2\left(\frac{\partial}{\partial X}+\frac{1}{\sqrt{\epsilon}}\Psi_X\frac{\partial}{\partial \xi}
+\frac{1}{\epsilon}\Theta_X\frac{\partial}{\partial \eta}\right)
\left((\gamma_0+\epsilon\gamma_1)\right.\\
& \quad \cdot\left.\left(\frac{\partial}{\partial X}+\frac{1}{\sqrt{\epsilon}}\Psi_X\frac{\partial}{\partial \xi}
+\frac{1}{\epsilon}\Theta_X\frac{\partial}{\partial \eta}\right)%\right.\\
\cdot
\left(\vphantom{\frac{\partial}{\partial X}}P_0+\epsilon P_1+\cdots\,,\right)\right)\\
& + \text{the same term with the $Y$-derivatives}\\
& + \left((\gamma_0+\epsilon\gamma_1)\left(P_{0z}+\epsilon P_{1z}+\cdots\,,\right)\right)_z\\
& + (\gamma_0+\epsilon\gamma_1)(n_0^2 + \epsilon\nu)\left(P_0+\epsilon P_1+\cdots\,,\right)=0\,.
\end{split}
\end{equation}
We put now
\begin{equation*}
\begin{split}
 P_0+\epsilon P_1+\cdots=
 (A_0(X,Y,z,\xi)+\epsilon A_1(X,Y,z,\xi)+\cdots)\E^{\I\eta}\,.
\end{split}
\end{equation*}
Using the Taylor expansion, we can formulate the interface conditions at $h_0$ which are equivalent
to interface conditions (\ref{InterfCond}) up to $O(\epsilon)$:
\begin{equation}\label{InterfCondh01}
\left(A_{0}+\epsilon h_1 A_{0z}+\epsilon A_{1}\right)_+ =(\text{the same terms})_-\,,
\end{equation}
\begin{equation}\label{InterfCondh02}
\begin{split}
& \left((\gamma_{0}+\epsilon h_1 \gamma_{0z}+\epsilon \gamma_{1})\right.\\
& \qquad\qquad\left.\times\left(A_{0z}+\epsilon h_1 A_{0zz}+\epsilon A_{1z}  - \epsilon\mathrm{i}\Theta_X h_{0X}A_{0}
 - \epsilon\mathrm{i}\Theta_Y h_{0Y}A_{0}\right)\right)_+ \\
& = \left(\mbox{the same terms}\right)_-\,.
\end{split}
\end{equation}

\subsection{The problem at $O(\epsilon^{0})$}
At $O(\epsilon^{0})$ we obtain
\begin{equation}\label{E0}
(\gamma_0 A_{0z})_z + \gamma_0 n^2_0 A_0 - \gamma_0\left((\Theta_X)^2 + (\Theta_Y)^2\right)A_0 = 0\,,
\end{equation}
with the interface conditions
$A_{0+} = A_{0-}$, $\left(\gamma_0 A_{0z}\right)_+ =\left(\gamma_0 A_{0z}\right)_-$
at $z=h_0$, and the boundary conditions $A_0=0$ at $z=0$ and $A_{0z}=0$ at $z= -H$.
We seek a solution to problem (\ref{E0}) in the form
\begin{equation} \label{anz0}
A_0 = B_j(X,Y,\xi)\phi(X,Y,z)\,.
\end{equation}
From eqs.~(\ref{E0}) we obtain the following spectral problem
for $\phi$ with the spectral parameter $k^2 = (\Theta_X)^2 + (\Theta_Y)^2$
\begin{equation} \label{Spectral}
\begin{split}
& \left(\gamma_0\phi_z\right)_z + \gamma_0 n_0^2 \phi - \gamma_0 k^2 \phi=0\,,\\
& \phi(0) = 0\,,
\quad \phi_z=0\quad \text{at}\quad z= -H\,,\\
& \phi_+ = \phi_-\,,\quad \left(\gamma_0 \phi_z\right)_+ =
\left(\gamma_0 \phi_z\right)_-\quad\mbox{at}\quad z=h_0\,.
\end{split}
\end{equation}
This spectral problem, considering in the Hilbert space $L_{2,\gamma_0}[-H,0]$ with the scalar product
\begin{equation} \label{L2scalar}
(\phi,\psi) = \int_{-H}^{\,0}\gamma_0 \phi\psi\,dz\,,
\end{equation}
has countably many solutions $(k_j^2,\phi_j)$, $j=1,2,\ldots$ where the eigenfunctions can be chosen as real functions.
The eigenvalues $k_j^2$ are real and have $-\infty$
as a single accumulation point.
The normalizing condition is
\begin{equation} \label{norm}
(\phi,\phi) = \int_{-H}^{\,0}\gamma_0 \phi^2\,dz=1\,,
\end{equation}

\subsection{The problem at $O(\epsilon^{1/2})$ and at $O(\epsilon^{1})$}
The solvability condition of problem at $O(\epsilon^{1/2})$ is
\begin{equation} \label{1/2}
\Theta_X\Psi_X+\Theta_Y\Psi_Y=0\,,
\end{equation}
from which we conclude that we can take $P_{1/2}=0$.

\subsection{The problem at $O(\epsilon^{1})$}
At $O(\epsilon^{1})$, we obtain
\begin{equation}\label{E1}
\begin{split}
&\left(\gamma_0A_{1z}\right)_z + \gamma_0 n_0^2 A_{1} - \gamma_0 k_j^2 A_{1} \\
&\quad = -\mathrm{i}\gamma_{0X}k_j A_0 -2\mathrm{i}\gamma_{0}k_j A_{0X}%\\
-\mathrm{i}\gamma_{0}k_{jX} u_0
+ \gamma_{1}k_j^2 A_0 - \gamma_{0}(\Psi_X)^2 A_{0\xi\xi} \\
& \qquad\quad\,-\text{the same terms with $Y$-derivatives}\\
& \qquad\quad\, - \frac{\partial}{\partial z}\left(\gamma_{1} A_{0z}\right) - n_0^2\gamma_{1}A_0 -\nu\gamma_{0}A_0 \,,
\end{split}
\end{equation}
with the boundary conditions $A_1=0$ at $z=0$, $A_{1z}=0$ at $z=-H$,
and the interface conditions at $z=h_0(X,Y)$:
\begin{equation}\label{InterfCondE1}
\begin{split}
& A_{1+}-A_{1-} = h_1(A_{0z-}-A_{0z+})\,, \\
& \gamma_{0+}A_{1z+}-\gamma_{0-}A_{1z-} \\
& \qquad = h_1\left(\left( (\gamma_0 A_{0z})_z\right)_-  -
\left( (\gamma_0 A_{0z})_z\right)_+\right) %\\
+ \gamma_{1-}A_{0z-}-\gamma_{1+}A_{0z+}\\
& \qquad -\mathrm{i}k_j h_{0X}A_0(\gamma_{0-}-\gamma_{0+})%\\
-\mathrm{i}k_j h_{0Y}A_0(\gamma_{0-}-\gamma_{0+})\,.
\end{split}
\end{equation}
Multiplying (\ref{E1}) by $\phi_j$ and then integrating resulting equation from
$-H$ to $0$ by parts twice with the use of interface conditions (\ref{InterfCondE1}), we obtain
the solvability condition for the problem at $O(\epsilon^1)$
\begin{equation}\label{MPE}
\begin{split}
&
2\mathrm{i}(\Theta_{jX} B_{jX}+\Theta_{jY} B_{jY})%\\
+ \mathrm{i}(\Theta_{jXX}+\Theta_{jYY}) B \\
& \qquad\qquad
+
((\Psi_X)^2+(\Psi_Y)^2) B_{j\xi\xi} + \alpha_j B_j = 0\,,
\end{split}
\end{equation}
where $A_0=B_j\phi_j$ and $\alpha_j$ is given by the following formula
\begin{equation*}
\begin{split}
&\alpha_j = \int_{-\infty}^0 \gamma_0\nu \phi^2_j\,dz
+ \int_{-\infty}^0 \gamma_1\left(n_0^2-k_j^2\right) \phi^2_j\,dz
 -
\int_{-\infty}^0 \gamma_1\left(\phi_{jz}\right)^2\,dz  \\
&  + \left\{h_1\phi_j\left[\left((\gamma_0\phi_{jz})_z\right)_+ - \left((\gamma_0\phi_{jz})_z\right)_- \right]
\vphantom{ \left(\frac{\gamma_1}{\gamma_0}\right)_-}
\right.\\
&\qquad\qquad\left.
\left.- h_1\gamma_0^2\left(\phi_{jz}\right)^2\left[\left(\frac{1}{\gamma_0}\right)_+ -
\left(\frac{1}{\gamma_0}\right)_-\right]\right\}\right|_{z=h_0}
\,.
\end{split}
\end{equation*}
Using spectral problem~(\ref{Spectral}), the interface terms introduced above
can be rewritten also as
\begin{equation*}
\begin{split}
&
\left\{h_1\phi_j^2\left[k_j^2\left(\gamma_{0+}-\gamma_{0-}\right)
- \left(n_0^2\gamma_0\right)_+ + \left(n_0^2\gamma_0\right)_-  \right]
\vphantom{ \left(\frac{\gamma_1}{\gamma_0}\right)_-}
\right.\\
& \qquad\qquad\left.
\left.- h_1\gamma_0^2\left(\phi_{jz}\right)^2\left[\left(\frac{1}{\gamma_0}\right)_+ -
\left(\frac{1}{\gamma_0}\right)_-\right]\right\}\right|_{z=h_0}
\,.
\end{split}
\end{equation*}

\section{The adiabatic mode Helmholtz equation and the ray parabolic equation in ray centered coordinates}
To obtain the adiabatic mode Helmholtz equation from eq.~(\ref{MPE}), we introduce the new amplitude
\[
D_j(x,y)=B_j(X,Y,\xi)\,,
\]
where $\displaystyle{(x,y)=\frac{1}{\epsilon}(X,Y)}$ are the initial (physical) coordinates.
One can easily obtain the following formulas for the $x$-derivatives of $D_j$:
\begin{equation}\label{d1x}
D_{jx}=B_{j\xi}\cdot \sqrt{\epsilon}\Psi_X+\epsilon B_{jX}\,,
\end{equation}
\begin{equation}\label{d2x}
D_{jxx}=B_{j\xi\xi}\cdot \epsilon(\Psi_X)^2 +
\epsilon^{3/2}(2B_{j\xi X}\Psi_X+B_{j\xi} \Psi_{XX})+
\epsilon^2B_{jXX}\,,
\end{equation}
and analogous formulas for the $y$-derivatives.
\par
The solvability condition of the problem at $O(\epsilon^{3/2})$ gives us
\begin{equation*}
\begin{split}
2B_{j\xi X}\Psi_X+B_{j\xi}\Psi_{XX} + 2B_{j\xi Y}\Psi_Y+B_{j\xi} \Psi_{YY}=0\,.
\end{split}
\end{equation*}
Substituting the obtained expressions for derivatives into eq.~(\ref{MPE}) we get,
after some manipulations, the reduced Helmholtz equation for $D$
\begin{equation}\label{rhe}
2\I(\theta_x D_{jx} + \theta_x D_{jy})+\I(\theta_{xx} + \theta_{yy})D_j%\\
+D_{jxx} +D_{jyy}+\bar\alpha_j D_j=0\,,
\end{equation}
where $\theta(x,y)=\frac{1}{\epsilon}\Theta(X,Y)$, $\bar\alpha_j=\epsilon\alpha_j$.
\par
This equation can be transformed to the usual Helmholtz equation
\begin{equation}\label{he}
\begin{split}
\bar D_{jxx} +\bar D_{jyy}+k^2 \bar D_j+\bar\alpha_j\bar D_j=0\,,
\end{split}
\end{equation}
where $k^2=(\theta_x)^2+(\theta_y)^2$ by the substitution $\bar D_j =D_j\exp(\I\theta)$.
Consider the ray equations for the Hamilton-Jacobi equation
\[
 (\theta_x)^2 + (\theta_y)^2={\cal P}^2+{\cal Q}^2=k^2
\]
 in the form
\begin{equation}\label{ray}
x_t =\frac{{\cal P}}{k}\,,
\quad y_t =\frac{{\cal Q}}{k}\,,\quad {\cal P}_t =k_x\,,
\quad {\cal Q}_t =k_y\,.
\end{equation}
We have $(x_t)^2+(y_t)^2=1$, so $t$ is a natural parameter for the ray, and
introduce $\vec{n}$ to be orthogonal to the ray (ray-centered coordinates).

To obtain the ray parabolic equation in the ray-centered coordinates, we first rewrite eq.~(\ref{he}) in the slow variables
 $(X,Y)=(\epsilon x,\epsilon y)$ (ray scaling)
\begin{equation}\label{hers}
\begin{split}
\epsilon^2\bar D_{jxx} +\epsilon^2\bar D_{jyy}+k^2 \bar D_j+\epsilon\alpha_j \bar D_j=0\,.
\end{split}
\end{equation}
Then, in the vicinity of a given ray, eq.~(\ref{hers}) can be written in the form
\begin{equation}\label{herc}
\begin{split}
\epsilon^2\frac{1}{h}(\frac{1}{h}\bar D_{jt})_t +\epsilon^2\frac{1}{h}(h\bar D_{jn})_n+k^2 \bar D_j+\epsilon\alpha_j \bar D_j=0\,,
\end{split}
\end{equation}
where $t$ is a natural parameter of the ray (arc length), $n$ is the (oriented) distance to the ray and
$\displaystyle{h=1-\frac{k_1}{k_0}}$.
  Hereafter we use, for a given function  $f=f(t,n)$, the following denotations:
$f_0=f|_{n=0}$, $f_1=f_n|_{n=0}$ and $f_2=f_{nn}|_{n=0}$.

Substituting into eq.~(\ref{herc}) the Taylor expansions
\begin{equation*}
k^2=k_0^2+2k_1k_0n+(k_1^2+k_0k_2)n^2%\\
=k_0^2+\sqrt{\epsilon}2k_1k_0N+\epsilon(k_1^2+k_0k_2)N^2\,,
\end{equation*}
\begin{equation*}
\frac{1}{h} =1+\frac{k_1}{k_0}n+\frac{k_1^2}{k_0^2}n^2
=1+\sqrt{\epsilon}\frac{k_1}{k_0}N+\epsilon\frac{k_1^2}{k_0^2}N^2\,,
\end{equation*}
\begin{equation*}
\frac{1}{h^2} =1+2\frac{k_1}{k_0}n+3\frac{k_1^2}{k_0^2}n^2
=1+\sqrt{\epsilon}2\frac{k_1}{k_0}N+\epsilon
3\frac{k_1^2}{k_0^2}N^2\,,
\end{equation*}
where $\displaystyle{N=\frac{1}{\sqrt{\epsilon}}}$ (parabolic
scaling), and the WKB-ansatz $\bar D_j=(u_0+\epsilon
u_1+\ldots)\exp((\I/\epsilon)\theta)$, we obtain at $O(\epsilon^0)$
\begin{equation*}
\begin{split}
\theta_t=\I k_0
\end{split}
\end{equation*}
and at $O(\epsilon^1)$ the parabolic equation in the ray centered coordinates
\begin{equation}\label{tayk1}
2\I k_0 u_{0t}+\I k_{0t}u_0+u_{0NN}+[(k_0k_2-2k_1^2)N^2+\alpha_{j0}]
u_0=0\,.
\end{equation}

\section{Mode Gaussian Beam Equations}

To solve eq.~(\ref{tayk1}), we first introduce the following
substitution:
\begin{equation}
u_0(t,N) =
\frac{1}{\sqrt{k_0(t)}}\exp\left(\frac{\I}{2}\int_0^t\frac{\alpha_{j0}(s)}{k_0(s)}\,ds\right)U_j(t,N)
\,.
\end{equation}
Then our equation becomes
\begin{equation}\label{tayk2}
2\I k_0 U_{jt}+U_{jNN}+(k_0k_2-2k_1^2)N^2U_j=0\,.
\end{equation}
Following~\cite{cerpop}, we seek a solution of this equation in the
form of the Gaussian beam anzats
\begin{equation}\label{gaussb}
U_j(t,N)=A(t)\exp\left(\frac{\I}{2}N^2\Gamma(t)\right)\,,
\end{equation}
where $\Gamma(t)$ is an unknown complex-valued function.
Substitution of (\ref{gaussb}) into (\ref{tayk2}) gives
$$
\I(2k_0A_t+A\Gamma)-AN^2[k_0\Gamma_t+\Gamma^2-(k_0k_2-2k_1^2)]=0\,.
$$
We require separately
\begin{equation}\label{gausseq}
k_0\Gamma_t+\Gamma^2-(k_0k_2-2k_1^2)=0\,,\quad\text{and}\quad
2k_0A_t+A\Gamma=0\,.
\end{equation}
To solve the first ordinary non-linear differential equation of the
Riccati type, we introduce new complex-valued variables $q(t)$ and
$p(t)$ by the formulas
$$
\Gamma = \frac{k_0}{q}q_t = \frac{p}{q}\,.%,\quad p = k_0 q_t\,.
$$
Then
\begin{equation}\label{gaussode}
q_t = k_0^{-1}p\,,\qquad p_t=(k_2-2k_1k_0^{-1})q\,.
\end{equation}
The solution of the second equation in (\ref{gausseq}) can be
expressed in the following form
$$
A(t)=\frac{\Psi}{\sqrt{q(t)}}\,,
$$
where $\Psi$ is a complex value, which is constant along the
ray, but may vary at different rays.

Finally for $u_0$ we have:
\begin{equation}
u_0(t,N) =
\frac{\Psi(\varphi)}{\sqrt{k_0(t)q(t)}}\exp\left(\frac{\I}{2}\int_0^t\frac{\alpha_{j0}(s)}{k_0(s)}\,ds
+\frac{\I}{2}N^2\frac{p(t)}{q(t)}\right)\,.
\end{equation}
Here $\varphi$ is the parameter, that enumerates rays. For $p$ and $q$ we have the system of ordinary
differential equations~(\ref{gaussode}), which can be solved simultaneously with the ray
equations~(\ref{ray}).
It is convenient to split variables $p$ and $q$ onto real and
imaginary parts as follows $p = p_2-\I\varepsilon p_1$, $q =
q_2-\I\varepsilon q_1$ where $\varepsilon$ is a rather big positive
real number, defining the width of the Gaussian beam. As found
in~\cite{cerpop} and discussed in~\cite{porter}, the optimal choice of $\varepsilon$ for the minimum
value of the Gaussian beam width for a homogeneous medium at the
point of the receiver corresponds to $\varepsilon=L_r$, where $L_r$ is
the length of the ray to the point of the receiver. Initial conditions
for $p$ and $q$ should be following
$$
q_1(0)=1\,,\quad p_1(0)=0\,,\quad q_2(0)=0\,,\quad p_2(0)=k_0(0)\,.
$$
The acoustic field at the point of the receiver $M$ can be expressed as
the integral on all rays
\begin{equation}\label{gbfield}
p(M) =
\int_0^{2\pi}\frac{\Psi(\varphi)}{\sqrt{k_0(t)q(t)}}\exp\left[\I\int_0^t\left(k_0(s)+\frac{\alpha_{j0}(s)}{2k_0(s)}\right)\,ds
+\frac{\I}{2}N^2\frac{p(t)}{q(t)}\right]\,d\varphi\,.
\end{equation}
Here $t$ and $N$ are the ray centered coordinates of the receiver
point for each ray.

One can determine the value of $\Psi(\varphi)$ by comparing of the
following two field. First, the one obtained for the homogeneous medium from the formula~(\ref{gbfield})
by the steepest descent method.
Second, the one obtained from the fundamental solution of the Helmholtz
equation for this case. So we have
$$
\Psi =\frac{\phi(z_s)\phi(z_r)}{\rho(z_s)}\cdot\sqrt{\I k_0(0)\varepsilon}\cdot
\sqrt{1-\frac{\alpha_{j0}(0)x}{2\I k_0(0)^2\varepsilon}\left(1+\frac{\alpha_{j0}(0)}{2k_0(0)^2}\right)^{-1}}
$$

\section{Numerical Example}

\begin{figure}[ptb]
\vspace{-2cm}
\begin{center}
\includegraphics[width=1.2\textwidth]{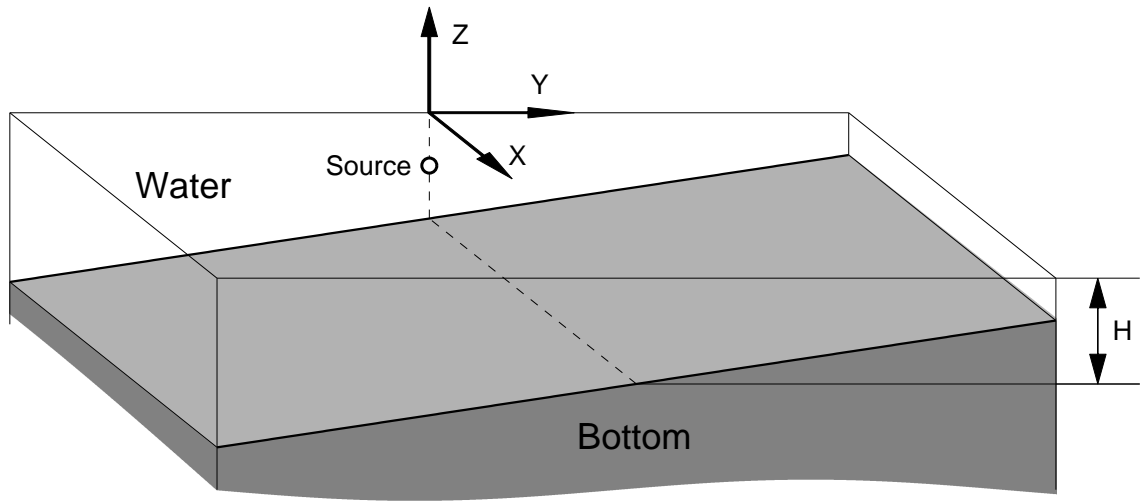}
\end{center}
\vspace{-1cm}
\caption{The geometry for the ASA wedge benchmark.
The wedge angle is $\approx 2.86^\circ$ with a distance to the apex $4\,km$.
The source is located at depth $100\,m.$ The bottom depth at the place of the source $H=200\,m$.}
\label{Fig1}
\end{figure}

\begin{figure}[ptb]
\begin{center}
\includegraphics[width=\textwidth]{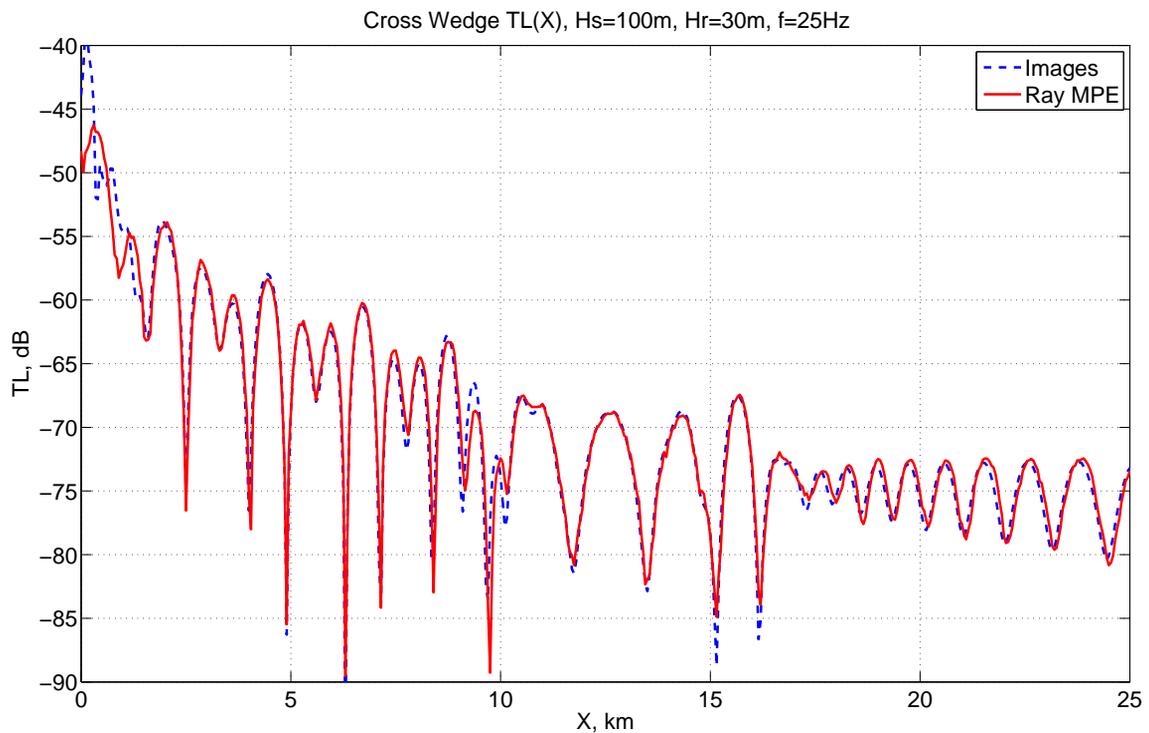}
\end{center}
\caption{The transmission loss for the ASA wedge, the source depth is $100\,m.$ The receiver depth is
$30\,m$, 3 modes, attenuation is $0.5\,dB/\lambda$. Across slope
propagation.} \label{Fig2}
\end{figure}

We consider a standard ASA wedge benchmark problem with the angle
of wedge $\approx 2.86^\circ$ in the case of cross slope propagation (see Fig.~\ref{Fig1}).
The bottom depth is $200\,m$ along the trace with $X=0\ldots 25\,km$.
The sound speed in the water is $1500\,m/sec$.
The sound speed in the bottom, which is considered liquid, is $1700\,m/sec$.
The bottom density is $1500\,kg/m^3$, the water density is $1000\,kg/m^3$.
We assume that there is no attenuation in the water layer, while in the bottom the attenuation
is $0.5\,dB/\lambda$. For calculation purposes we restrict the total depth to
$600\,m$.

To illustrate the efficiency of our equation, we performed a
numerical simulation of sound propagation for the standard ASA wedge
benchmark.
In fig. \ref{Fig2}, we present comparisons of the solution of
our equation and the source images solution \cite{buck} in the case of
cross slope propagation in the wedge with ASA parameters. One can see that the
curves are quite close, and the mean square difference between curves is about $1.6\,dB$
in the case of 3 modes. To improve accuracy of the method on the first $1.5\,km$ we can use
more than 3 modes. For example, in the case of 7 modes, the field in the vicinity of the source
is represented correctly, and the mean square difference is about $1.4\,dB$.

\section{Conclusions}
The results of test calculations show, that the acoustic field in the far zone is
satisfactory described by its first three modes.
We have shown that no interaction of modes is necessary
to perform satisfactory modeling of a cross slope propagation.
However, to obtain a more realistic model, we assert, that
seven modes (total depth is $600\,m$) are sufficient to represent the acoustic
field in the all considered area.

\section*{Acknowledgements}
The authors are grateful for the support to ``Exxon Neftegas Limited'' company.

\end{document}